Web 地図画像の利用と著作権法

岩崎亘典\*・飯田　哲\*\*

Use of "Web Map Image" and copyright act

Nobusuke IWASAKI\*, Satoshi IIDA\*\*
\*　　正会員　農研機構農業環境変動研究センター（Institute for Agro-Environmental Sciences, NARO）
　　　〒305-8604　茨城県つくば市観音台３－１－３　E-mail: niwasaki@affrc.go.jp
\*\*　非会員　合同会社 Georepublic Japan

In this paper, we reviewed the notes on using Web map image provided by Web map service, from the viewpoint of copyright act. The copyright act aims to contribute to creation of culture by protecting the rights of authors and others, and promoting fair exploitation of cultural products. Therefore, everyone can use copyrighted materials to the extent of the copyright limitation based on copyright act. The Web map image, including maps, aerial photo and satellite image, are one of copyrighted materials, so it can be used within the limits of copyright. However, the available range of Web map image under the copyright act is not wide. In addition, it is pointed out that the copyright act has not been able to follow the progress of digitalization of copyrighted materials. It is expected to revise the copyright act corresponding to digitalization of copyrighted work.

Keywords: Web 地図画像（Web Map Image），著作権法（Copyright Act），著作権の制限（Limit of Copyright）

## １．はじめに

　Web の普及により様々な情報の利便性が高くなった．中でも地図の利用は大きく影響を受けたものの一つである．特に近年ではスマートフォンの爆発的な普及にともない，Web 地図サービスや地図アプリなどは日々の生活に必須のものとなっている．研究や行政における GIS の利用や研究においても，WebGIS の登場と一般化は大きな変化をもたらしている．さらに，地理空間に関する各種情報のオープンデータ化，FOSS4G[1]の普及，地図タイル形式による効率的なデータ配信方式の確立により，日本の国土全体程度であっても，個人が地図情報を提供することが可能となった[2)3)]．

　この様に，Web で公開されている地図データやサービスは様々な恩恵をもたらした一方で，

簡便に利用できるようになったために，生じる問題もある．その代表が著作権侵害であり，2016〜2017 年にかけて，各地の地方自治体において著作権侵害の恐れがある地図画像をウェページ上に公開していたことが問題となった [4)5)6)]．

これまで地図および地理空間情報と著作権の関係を扱ったものとしては，主に行政機関が有する地理空間情報の円滑な活用のためのガイドラインの作成（地理空間情報活用推進会議，2010；測量行政懇談会，2011）や，井上（2013）による地図の著作物の創作性に関する考察などがある．しかしこれらは，一般の利用者による Web 地図サービスを対象としたものではない．また，Web 地図サービスについて論じたものとしては，利用規約に基づく適正な利用法についての解説があるが [7)8)]，著作権そのものについて言及したものではない．しかし，Web 地図サービスのみならず，Web 上の様々な著作物を適切に利用するためには，著作権法が何を，どこまで保護するのかについて理解することが必要である．

さらに近年では，地理空間情報に限らず様々な情報がオープンデータとした公開されているが，これらの情報を適切に利用することや，自らが有する情報をオープンなライセンスを付与して公開するにあたっても，著作権に関する理解が必要となる．

そこで本稿ではいわゆる Web 地図サービスで提供されている地図画像を利用する場合，どのような様態であれば著作権を侵害することなく利用することが可能であるのかについて解説する．そのためにまず，著作権について概説をおこなうとともに，Web 地図サービスで多く見られる利用許諾と著作権の関係について整理する．その上で，著作権の範囲の中での地図画像の利用可能性について検討するとともに，我が国における地図利用にあたって無視することができない測量法についても言及する．

なお本稿では，地図データ，Web 地図サービス，地図画像を以下の通り定義する．
- 地図データ：デジタル化された地理空間情報
- Web 地図サービス：Web を介して提供される位置検索，経路検索，地図閲覧等のサービス
- 地図画像：Web 地図サービスにより提供される画像情報としての地図や空中写真，衛星画像

また，本論文の著者らは法律家ではない．そのため，本論文の内容も，法律の解釈ではなく，実際の利用に当たっての留意点についてであることをご理解頂きたい．

## ２．著作権の概要

著作権および著作権法については，多くの書物や専門家により論じられている．また，特に近年の Web やデジタル技術の進展に伴う著作権の変質といった視点では，野口（2010）や水野（2017）などがある．著作権に関わる詳細についてはこれらの書物を参考いただくとし，ここでは論説を進めるにあたり必要となる事項について，簡単にまとめる．

### ２．１．著作権の特徴とその対象

まず著作権とは，創作された著作物について，その創作者である著作者のみがとりうる利用方法について，著作権法で規定されている権利である．我が国の著作権法では，目的を記した第 1

条に「文化的所産の公正な利用に留意しつつ、著作者等の権利の保護を図り、もつて文化の発展に寄与することを目的とする」とある．すなわち，著作権法の基本的考え方は，権利の保護と利用の促進の両方を通じて，文化の発展を目指すことである．

表1に著作権法の保護対象と対象とならないものについて示す．著作権で保護される著作物については，第2条1項1号に「思想又は感情を創作的に表現したものであつて、文芸、学術、美術又は音楽の範囲に属するものをいう」とされており，第10条に代表例が例示されている．ここで重要なのは，「思想又は感情を創作的に表現したもの」とする点で，誰が作成しても同じような表現になる簡単な図表、誰が観測しても同じになる事実データ、そして事実データを単純に集積したデータベース（DB）等は「創作的に表現したもの」に該当しないため著作物とは認められない．ただし，DBでその情報の選択又は体系的な構成によって創作性を有するものは，著作物として保護される（第12条2項）．

また，著作権法の特徴として権利の取得に申請が必要ない無方式主義を取っている事があげられる（第17条2項）．そして著作権法により保護されている具体的な内容は，第18条から第28条に記載されている各種の権利である．ここに記載された行為については，著作権者または著作者の許諾を得ることなしに行うことができない．つまり著作権は，どのような著作物であっても自動的に発生し，多くの行為が著作権者の許諾を必要とすることから，様々な知的財産権の中でも極めて強い権利である．そのため，第1条にある権利の保護と利用の促進のバランスをとる観点から保護期間が定められており，その期間を終えた著作物は自由に利用することが可能となる（第51～58条）．また，法律など広く利用されることが求められるものについては，著作権による保護は認められない（第13条）．

2．2．著作権による保護と利用許諾の関係

さて，著作権法による保護には人格的な権利を保護する著作者人格権（第18条から第20条）と，財産権として著作財産権（第21条から第28条）の二つがある．Web地図画像の利用に大きく関係する財産権としての著作財産権について概説する．

Web地図画像の利用に関する著作権の保護としては，第21条複製権，第23条公衆送信権等，第27条翻訳権、翻案権等があげられる．つまり，地図画像を閲覧することは妨げら

表1 著作法による保護対象と対象ではないもの

| 保護対象 | 保護対象でないもの | |
|---|---|---|
| 著作物 | 著作物 | 著作物でないもの |
| ・小説、脚本、論文、講演その他の言語の著作物<br>・音楽の著作物<br>・舞踊又は無言劇の著作物<br>・絵画、版画、彫刻その他の美術の著作物<br>・建築の著作物<br>・地図又は学術的な性質を有する図面、図表、模型その他の図形の著作物<br>・映画の著作物<br>・写真の著作物<br>・プログラムの著作物 | ・保護期間が終わったもの<br>・法律等<br>・判決 | ・事実の伝達にすぎない雑報及び時事の報道<br>・データベース<br>・事実データ<br>・アイディア<br>・タイプフェイス<br>・人や物のパブリシティ権 |

れないが，閲覧したものを複製（印刷）したり、他者が閲覧可能な形で公開したり，情報の抽出や形式の変換などの翻案等を行うことは，著作権者の許諾を得ること無しに行うことができない．

ただし、前述の通り「公正な利用」を促進するために，「第2章第3節第5款　著作権の制限」として、著作権者に許諾を取らなくても利用可能な場合が規定されている。例えば私的使用がその代表例であるが，研究や業務等で地図や地理空間情報を使う場合は私的使用に該当するとはいえない．そのため本来であれば，こうした作業を行うには著作権者から利用許諾を得る必要があるが，その様な利用許諾を一々得ることは現実的に困難であるといえる．

そのため、利用許諾契約または使用許諾契約を定めることにより、申請を介することなく一定条件の下で利用許諾を与えることが行われている。World View2 等の高解像度衛星画像を提供している Digital Globe 社は利用許諾契約例として「Internal Use License」[9] を提示しており，そこに記載されている範囲であれば，事前に利用を許諾している．

今一度整理するが，利用許諾契約は著作権によって使用が制限される事項について，許諾を与えるものである．つまり，地図画像の利用を考える場合，はじめに著作権の範囲の中での利用が可能であるか，次に各種 Web 地図サービスの利用許諾の範囲での利用可能であるかを考慮する必要がある．

なお本稿は，主に著作権の観点から地図画像の利用を考えるという趣旨である。利用許諾契約に関しては，それぞれの Web サービスの利用許諾契約がサービスごとに異なるため本稿では詳細には言及しない．しかし，利用許諾契約に明示的に許可されていない使用については，許諾者と利用者の解釈が異なる可能性があるため，原則的に避けることが望ましいと考えられる．

## ３．地図画像の著作性と使用上の留意点

Web 地図サービスでは，住所や経路，POI 情報検索などの様々な情報が提供されている。これらのうち，本論では特に地図画像として，地図，衛星画像または空中写真画像について解説することとする。まず，これらの情報が著作物に該当するのかどうかを検討し，その上で我が国の著作権法の範囲で許諾なく使用できる範囲について検討する。

### ３．１．地図画像の著作物性

まず地図全般については，表1に示した著作権法の例示にも出ていることから，基本的に著作物として認められる。しかしすべての地図が著作物として認められるわけではなく，ごくありふれた手法により描かれた略図であれば著作物として認められない判例が存在する（東京地方裁判所昭和54年6月20日判決（昭和54(ワ)1341））。すなわち地図に著作物性が認められるのは，地図に描画されている対象の取捨選択や，凡例や彩色，注記等の位置やフォントの選択といった描画方法に著作物性が認められ，著作物として保護されることとなる．

この観点で見た場合，井上（2013）は作成マニュアルが厳密に定められている場合につ

いては，同様の手順において地図を作成した場合，作業者の著作権侵害は否定されるだろうとしている．また，地理空間情報活用推進会議（2010）も「測量成果としての地図の作成作業において、作業者が創作性を発揮する余地は限定される」としている．

一方でこれは，作成マニュアルが厳密に整備されている場合や公共測量に限ったことである．各 Web 地図サービスで提供される地図画像は，それぞれのサービスの提供主体が独自の描画方法を採用している．また，Web 地図サービスで提供される地図画像ではないが，Web 上で公開されている地図について著作物性が認められ再配布が禁じられた判例もある（京都地方裁判所平成 13 年 5 月 31 日判決（平成 10(ワ)3435））．以上のことから，Web 地図サービスで提供されている地図画像については，著作物として認めることが妥当だと考えられる．

空中写真については，地理空間情報活用推進会議（2010）は，公共測量として実施する空中写真撮影は作業規程が細かく定められており，作業規定に沿って空中写真を調製する場合は創作性を発揮できる余地が少ないため，著作物とは認められない可能性があるとしている．一方で，地図の場合と同様にこれらのマニュアルに準拠しない空中写真や衛星画像については，撮影する高度や撮影範囲の決定，人工衛星の場合は撮影におこなうセンサーのバンドの選定など，撮影者によって独自に決定される余地がある。そのため、これらの点に創作性を認めることは可能であり，著作物として認めることが妥当であろう．

３．２．地図画像の使用について

３．１．で示したとおり，Web 地図サービスで公開されている地図画像は，基本的に著作物と見なすことが妥当であると考えられる．すなわち，使用に当たっては著作権法で規定されている範囲であれば，自由に利用することが可能である．これは，２．２．でも示したが，著作権法の著作権の制限に該当する利用であり，詳細な例示については，文化庁の Web サイト [9]にも記載されている．中でも一般の利用にあたって関係の深いものは、第 30 条「私的使用のための複製」、第 32 条「引用」であろう．加えて教育機関においては第 35 条「学校その他の教育機関における複製等」と第 36 条「試験問題としての複製等」も地図の利用にあたっては，深い関係があるだろう．以下、上記の四条に関連して地図画像がどこまで利用可能であるか検討する．

３．２．１．私的使用のための複製

私的使用については，「個人的に又は家庭内その他これに準ずる限られた範囲内」において複製が許可されている．ここでは，「家庭内その他これに準ずる限られた範囲内」をどこまでととらえるかにより，使用可能な範囲が決定される．地図に関連した報告ではないが，昭和 56 年著作権審議会第 5 小委員会報告書 [10]によれば「親密な特定少数の友人間、小研究グループなどについては、この限られた範囲内と考えられるが、少人数のグループであってもその構成員の変更が自由であるときには、その範囲内とはいえないものと考える」としている．以上から私的使用のための複製は，あくまで個人か極めて限定された少人数の間でのみ可能だと考えられる．また，個人や少人数であっても，例えば社内会議上での配

布など業務上使用するために著作物を複製することは私的使用とは見なされないので注意が必要である．

３．２．２．引用

上記の私的使用に比べて，引用に基づく使用であれば，一般に公開する目的であっても利用可能である（第32条）．引用の要件として文化庁のWebページ[11)]では，

ア．既に公表されている著作物であること

イ．「公正な慣行」に合致すること

ウ．報道、批評、研究などのための「正当な範囲内」であること

エ．引用部分とそれ以外の部分の「主従関係」が明確であること

オ．カギ括弧などにより「引用部分」が明確になっていること

カ．引用を行う「必然性」があること

キ．「出所の明示」が必要（コピー以外はその慣行があるとき）

をあげている．地図画像を「引用」して利用する様態として考えられるのは，例えば地図画像自体の批評や比較がある．例えば二つの地図を比較し，どちらの地図の視認性が高い，情報量が多い，といった点を論じるのであれば，主となる論点は視認性や情報量であり，地図画像は比較材料として用いられる従であるといえる。人工衛星画像の場合も同様で，衛星画像の解像度や使用された波長の違いが，データとしての特性に及ぼす影響を論じるのであれば，主はデータの特性であり，衛星画像自体は従であるといえる．

なお，引用についての例示の多くが文章についてであり，地図画像の使用を考えるのに難しい側面があった．しかし，「脱ゴーマニズム宣言事件」として知られる判決（東京高等裁判所平成12年4月25日判決（平成11(ネ)4783））では，「意見を批評，批判，反論するために，その意見を正確に指摘しようとすれば，漫画のカットを引用することにならざるを得ない」として，漫画のカットの引用が認められた．地図画像についても同様と考えられ，必要性があれば引用することは可能であると考えられる．

しかし前述のア〜キの要件を満たさない場合，引用とするのは難しいと考えられる．例えば，案内図を作成するために地図画像をキャプチャーし，情報を付加した上で公開されている事例が散見される．また学会発表等でも，背景画像として地図画像が利用される事例が散見される．こうした事例では，背景として使用している地図画像自体が批評や研究の対象ではなく，そのために付加情報が主，背景画像が従であるとは言い難く，引用とするのは困難である．また，衛星画像等から地物の抽出やトレース等を行い公開する事例も見受けられるが，こうした事例では画像自体に手を加えており，引用には該当しない．

これらの行為は，著作権法上は同一性保持権（第20条），複製権（第21条）、翻案権等（第27条）の侵害にあたると考えられる．同時に，その成果をWeb上に公表することや，報告書や論文として公表することは，公衆送信権等（第23条）や譲渡権（26条の2）の侵害であると考えられる．

３．２．３．学校その他の教育機関における複製等

2022年度以降の高校教育課程での地理総合の必修化や，その中でGIS教育が明記されていることから，学校教育の現場において，これまで以上に多くのWeb地図サービスが利用されることが予想される．しかし，著作権法上認められているのは，「必要と認められる限度において、公表された著作物を複製する」（第35条）ことであり，複製以外は許諾が必要であることに留意が必要である．またこの条文の対象となるのは「学校その他の教育機関（営利を目的として設置されているものを除く。）において教育を担任する者及び授業を受ける者」であることから，例えば大学やその他研究機関において、授業ではなく研究目的で地図画像を複製することは含まれていないことにも注意が必要である．

３．３．公共測量により作成された地図の使用について

　ここまで主に，地図画像を著作物と見なした場合に，著作権法の範囲内でどの様な利用が可能であるかについて論じた．しかし，日本国内において刊行されている地図のうち，基本測量および公共測量の測量成果については，測量法によっても使用が制限されている．Web地図サービスで提供される地図画像としては，地理院タイルのうち基本測量成果に該当するものが対象となる．これらについては，測量法に承認申請が必要な場合があるが，一方で承認を得ず出所の明示により利用できる範囲も規定されている．この範囲には私的に利用する場合，刊行物等に少量の地図を挿入する場合，学校その他教育機関で利用する場合，学術論文に利用する場合等が例示されおり，３．２．１〜３までにあげた利用方法であれば，出所の明示を条件とした利用が可能である[12]．

## ４．まとめにかえて　〜著作権法の将来への希望

　以上のように，Web地図サービスで公開されている地図画像については，私的使用，引用等，著作権法で許可された範囲での利用が可能である．ここでまとめた著作権法の基礎を，地図画像利用の参考になることを期待する．なお実際には，利用許諾契約も確認することが必要であるが，そこに明示されてない使用は避けるべきだろう．

　以上のように，著作権法に基づく地図画像の利用は，制限が大きいといわざるをえない．一方で，近年のデジタル技術の進展に対して，著作権法が追いついていないという指摘も多くある（水野，2017；田中，2017）．こうしたなか，著作権法の存在を前提としつつ，より自由な利用を可能とするCreative Commons[13]のような利用許諾契約もつくられている．また，Web地図サービスとしても，OpenStreetMap[14]の様に一定の許諾条件下で自由に利用可能なもの存在する．

　こうした利用者側からの取り組みも必要だが，著作権法についても，現在の状況に応じた改正等も期待されるだろう．また，初めにあげた地方自治体による地図画像使用違反では，測量法の規定に違反した事例が多く見られた．しかし，地図画像の利用が萎縮や自粛して行政サービスが低下することになれば，本末転倒といえるだろう．著作権法も測量法も，その基本精神は，健全な範囲での保護と利用を両立することであり，その趣旨を果たすために法律やそれに準ずる規程も変わっていく必要があるだろう．

官民データ促進法の成立や，それを考慮して改訂された「世界最先端 IT 国家創造宣言・官民データ活用推進基本計画」により，現在，様々なデータ利用のさらなる促進も求められている．本稿がそうした利用に当たっての参考になるとともに，地図画像の利用に関しても，適切な範囲での利用可能性が拡大することを期待したい．

## 謝辞

## 注

1) Free and Open Source Software の頭文字を繋げたもので，オープンソースの地理空間ソフトウェアの総称
2) http://www.gridscapes.net/
3) http://tiles.dammaps.jp/ryuiki/
4) http://www3.pref.nara.jp/hodo/item/54860.htm
5) http://www.pref.kagawa.lg.jp/content/koho/houdou/wzcr4i161115164901.shtml
6) https://www.pref.miyagi.jp/release/ho20170330-2.html
7) https://qiita.com/nyampire/items/5fd06107f25bc12a526f
8) http://koutochas.seesaa.net/article/444993828.html
9) https://www.digitalglobe.com/legal/internal-use-license
10) http://www.cric.or.jp/db/report/s56_6/s56_6_main.html
11) http://www.bunka.go.jp/chosakuken/naruhodo/outline/8.h.html
12) http://www.gsi.go.jp/LAW/2930-qa.html
13) https://creativecommons.org/
14) http://www.openstreetmap.org/

## 参考文献